\documentclass{llncs}
\input epsf
\usepackage{makeidx}
\usepackage{amsmath}
\begin{document}
\pagestyle{empty}
\title{Lattice structure and convergence of a Game of Cards
\thanks{This work have been done during the time of M.M and H.D.P in 
Departamento de Ingenier\'{\i}a 
Matem\'atica, Universidad de Chile and was supported by Project ECOS-C96E02
and Chilean program FONDAP in Applied Mathematics (E.G., M.M., H.D.P.)
}}
\author{Eric Goles\inst{1} \and Michel Morvan\inst{2} \and Ha Duong Phan\inst{3}}
\institute{Departamento de Ingenier\'{\i}a Matem\'atica,
Escuela de Ingenier\'{\i}a, Universidad de Chile, Casilla 170-Correo 3,
 Santiago, Chile,\\
\and
LIAFA Universit\'{e} Denis Diderot 
Paris 7 and Institut universitaire de France - Case 7014-2, Place Jussieu-75256 Paris Cedex 05-France,\\ 
\and
LIAFA Universit\'{e} Denis Diderot 
Paris 7 - Case 7014-2, Place Jussieu-75256 Paris Cedex 05-France,\\
\email{egoles@dim.uchile.cl, morvan@liafa.jussieu.fr, phan@liafa.jussieu.fr}  
}
\maketitle              
\begin{abstract} 
 We study the dynamics the so-called Game of Cards
  by using tools developed in the context of discrete dynamical
  systems. We extend a result of~\cite{Des95} and of ~\cite{Hua93}
  (this last one in the context of distributed systems) who established a
  necessary and sufficient condition for the game to converge. We
  precisely describe the structure of the set of configurations (that
  we show to be very closed to a lattice structure) and we state bounds
  for the convergence time.

\end{abstract}

\section{Introduction}

This paper is devoted to the study of the dynamics of a discrete
system related to some self stabilizing protocol on a ring of
processors. As explained in~\cite{Des95}, this protocol can be seen in
terms of a game of cards described as follows. ``Assume a finite set
of players sitting around a table. Initially, each player holds a
finite number of non-distinguishable cards. The only move a player can
make is passing a card to his/her right neighbor, provided that this
neighbor has fewer cards than the player itself. The game terminates
when no move is possible.'' In the cited paper, the following theorem
is proved.

\begin{theorem}
\label{Desel}
The Game of Cards terminates if the total number of cards is a
multiple of the number of players.
\end{theorem}

The proof given by the authors for this theorem was simpler than the
one proposed for the equivalent result in~\cite{Hua93}. Moreover, the authors
werw pointing out the fact that studying some distributed protocols in
terms of discrete dynamical systems could be fruitful. In this paper,
we replace the game of cards in the broader context of the study of
transition systems on compositions of a given integer, where the
total number of cards which is decomposed in each configuration is the
sum of the number of cards of each player. The dynamics of such
transition systems on compositions has been intensively studied by
various authors and provides a powerful framework to derive structural
and dynamical properties~\cite{Bry73,GK93,GK86,GMP97,GMP00}. 

In this paper, we are going to investigate in more details the structure
 of the set
of all possible configurations of the game with $n$ cards and $p$
players. For that, let us represent a configuration by a list of $p$
integers $a=(a_1,\ldots,a_p)$ where $a_i$ is the number of
cards of player $i$. At each step player $i$ can give a card to
 player $(i + 1)$ (modulo $p$). Let $n= kp+q$ with $0
\leq q < p$. Let us call $\mathcal G$ the graph defibed on the set
of all possible configurations of the game --- that is the set of
$p$-dimensional vectors of integers such that the sum of the
components is $n$, and having for arcs the set of couples
$(a,b)$ of configurations such that $b$ can be obtained from $a$ in
one step. Following \cite{Des95}, let us call dual those configuration
that do not belong to a non trivial circuit.
The Game of Cards can be coded by a Chip Firing Game of which many propeties have been studied in \cite{LP00}. Moreover, due to its special rule, the Game of Cards has other interesting propeties that we will studying in this paper.
 We will first characterize dual configuration
and will show that if $n$ is not a multiple of $p$, the unique non
trivial strongly connected component of the graph $\mathcal G$ is the
set of dual configurations. We will also study the subgraph of $\mathcal
G$ induced by the set of configurations that can be reached from a
given configuration $O$. We will characterize the partial order
naturally associated to this graph when the set of dual configurations
is replaced by a unique vertex, and we will establish its lattice
structure. We will finish by bounding the number of steps
necessary to arrive to a dual configuration.

\bigskip

In the following, we are going to discuss some lattice
properties of the above dynamical systems. Let us recall that a finite
lattice can be described as a finite partial order such that any two
elements $a$ and $b$ admit a least upper bound (denoted by $sup(a,b)$)
and a greatest lower bound (denoted by $inf(a,b)$). $Sup(a,b)$ is the
smallest element among the elements greater than both $a$ and $b$.
$Inf(a,b)$ is defined similarly. A useful result about finite lattices
is that a partial order is a lattice if and only if it admits a
greatest element and any two elements admit a greatest lower bound.
For more informations about lattice theory, see \cite{Bir67,Dav90}.

\section{Basic structure of the Game of Cards}

Let us first state the following corollary of the main theorem of \cite{Des95}.

\begin{corollary}
  
  If $q = 0$ there is no dual configuration, which means that the game
  terminates; if $q>0$, the game does not terminates and the dual
  configurations are exactly the $q \choose p$ 
  configurations such that each player
  owns either $k$ or $k + 1$ cards.

\label{corollary:Desel}
\end{corollary}
\begin{proof}   
The proof comes immediately from the proof of \cite{Des95}

\qed  \end{proof}

See Figure 1 for two examples:

\begin{figure}[h]
$${\epsfbox{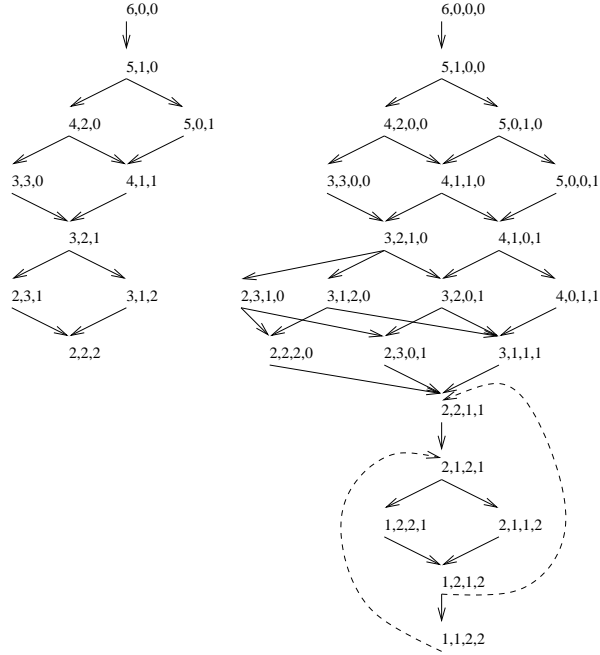}}$$
\caption{The Game of Cards in two cases: 6 cards and 3 players, and 6 cards and 4 players. $ a \longrightarrow b$ signifies that $b$ can obtained from $a$ by moving a card from a player positioned between 1 and $p-1$. $ a --> b$ signifies that $b$ can obtained from $a$ by moving a card from the $p-th$ player to the first player.}
\label{fig:01}
\end{figure}

We can now state the following theorem.

\begin{theorem}
\label{theo:connexe}
  
  The unique non trivial strongly connected component of $\mathcal G$
  is the set of dual configurations.

\end{theorem} 
\begin{proof}  

The result is obvious if $q = 0$. If $q > 0$, it is clear from the
definition that a non dual configuration can not belong to a strongly
connected component since it does not belong to a circuit. So we just
have to prove that the set of dual configurations is a strongly
connected component.

Let $a=(a_1,\ldots,a_p)$ and $b = (b_1,\ldots b_p)$ be two dual
configurations. We have $\forall i \in [1,p]$, the value of 
$a_i$ and $b_i$ is $k$ or $k+1$. Let $P$ be the dual configuration
defined by 
$$P=\underbrace{k+1,\ldots,k+1}_{q}, \underbrace{k,\ldots,k}_{p-q}.$$
We are going to show that there exists a path from $a$ to $P$ and a
path from $P$ to $b$, which implies the existence of a path from $a$
to $b$. Similarly, the existence of a path from $b$ to $a$ could be
stated.

The path from $a$ to $P$ is built by following an arbitrary maximal
path starting from $a$ and in which no transition is made in which
player $q$ gives a card to its right neighbor (such a path exists
since in each infinite path player $q$ plays an
infinite number of times). The unique possible transition from the
last configuration of this path is the one in which player $q$ plays,
which proves that this configuration is $P$.

We have now to find a path from $P$ to $b$. Let $i$ be the last index
smaller than or equal to $p$ such that $b_i =k+1$. By consecutively
applying the playing rule from configuration $P$ on players $q$,
$q+1$, $\ldots$, $i-1$, we obtain the following configuration:
$$
\underbrace{k+1,\ldots,k+1}_{q-1},
k,\ldots,k,\underbrace{k+1}_{i th},k,\ldots,k.$$
Let $j$ be the last index smaller than $i$ such that $b_j = k+1$. We
can apply the same techniques and by iterating the process, we obtain
the configuration $b$. This achieves the proof.
\qed  \end{proof}

If we remark now that when $q > 0$ there is no fixed point, the
previous result allows to state:

\begin{corollary}
There is a path from any configuration to any dual configuration.
\end{corollary}

If $q > 0$, any infinite sequence of transitions starts by a finite
sequence of transitions applied to a non dual configuration and leading
to another non dual configuration and is followed by an infinite
sequence of transitions applied to dual configurations and leading to
dual configurations. In some sense, these dual configurations
represent a generalization of fixed points, since when the system is
in such configuration it cannot reach a non dual configuration but it
can reach any dual one. So it
is natural to consider the reduced graph $R({\mathcal G})$ obtained
from ${\mathcal G}$ by replacing the set of dual configurations by a
unique vertex $\bot$ with no outgoing arc and with one ingoing arc
coming from each non dual configuration that can lead in one
step to a dual configuration. ($R(\mathcal G)$ is then the
quotient graph relatively to the equivalence relation ``is in the same
strongly connected component of''). We can now consider the partial
order $<_{gc}$ on $R(\mathcal G)$ by considering the
 transitive closure of the graph: $b <_{gc} a$ if $b$ is reachable
from $a$ by following a path in the graph.

We are now going to focus on the relation between non dual
configurations by first showing that between
two such configurations the length of all paths is the same. For that,
we are going to introduce the notion of ``shot vector'', following 
\cite{Eri93},  which describes for each player the number of cards it 
gave to its neighbor.

Let $a$ and $b$ be two non dual configurations 
such that $b <_{gc} a$.
Let $\mathcal{C}$ be a sequence of transitions from $a$ to $b$:
$${\mathcal C} : a \rightarrow c^1 \rightarrow c^2 \rightarrow \ldots
\rightarrow c^l \rightarrow b.$$
Let $s_i({\mathcal C})$ be the number of cards given by player $i$ 
to its neighbor during this sequence. Let $s({\mathcal C})$
be the sequence $(s_1({\mathcal C}),s_2({\mathcal
  C}),\ldots,s_p({\mathcal C}))$ and call it the {\em shot vector} of the
sequence ${\mathcal C}$. Let $|s({\mathcal C})| = \sum_{i=1}^{i=p}
s_i({\mathcal C})$ be the {\em length} of the sequence ${\mathcal C}$. 
It immediately comes that:
$$s_2({\mathcal C}) -s_1({\mathcal C}) =  a_2 - b_2,$$
$$s_3({\mathcal C}) -s_2({\mathcal C}) =  a_3 - b_3,$$
$$\vdots$$
$$s_1({\mathcal C}) -s_p({\mathcal C}) =  a_1- b_1,$$
which implies that 
$$s({\mathcal C}) = (s_p({\mathcal C}), \ldots,s_p({\mathcal C})) +
(a_1-b_1, a_1+a_2 -b_1-b_2,\ldots,a_1+\ldots
+a_{p-1}-b_1-\ldots-b_{p-1},0).$$
If we denote $$d(a,b)=(a_1-b_1, a_1+a_2
-b_1-b_2,\ldots,a_1+\ldots +a_{p-1}-b_1-\ldots-b_{p-1},0),$$
we obtain
$$s({\mathcal C}) = s_p({\mathcal C}) *(1,\ldots,1) + d(a,b).$$
Let us now introduce the following partial order between shot vectors: 
given two shot vectors $s({\mathcal C})$
and $s({\mathcal D})$, $s({\mathcal C}) \leq s({\mathcal D})$ if
$\forall i$ $s_i({\mathcal C}) \leq s_i({\mathcal D})$. Let $a$ and
$b$ be two elements reachable from a given configuration $O$.  Let us
consider two sequences of transitions, one from $O$ to $a$, the other
from $O$ to $b$:
$${\mathcal C} : O \rightarrow c^1 \rightarrow c^2 \rightarrow \ldots
\rightarrow c^u \rightarrow a,$$
$${\mathcal D} : O \rightarrow d^1 \rightarrow d^2 \rightarrow \ldots
\rightarrow d^v \rightarrow b.$$
We can state the following lemma:

\begin{lemma}

Assume that there exists an index $j$ such that $s_j({\mathcal C}) \leq
s_j({\mathcal D})$ and $\forall j' \neq j, \quad s_{j'}({\mathcal C})
\geq s_{j'}({\mathcal D})$. If it is possible to apply the rule on
position $j$ of $b$, then it is also possible to apply this rule to
$a$ in the same position $j$.

\label{lemmes:pos}
\end{lemma}
\begin{proof}  

From the shot vector definition, we obtain:
$$a_1 = O_1 -s_1({\mathcal C}) + s_p({\mathcal C}),$$
$$a_2 = O_2 -s_2({\mathcal C}) + s_1({\mathcal C}),$$
$$\vdots$$
$$a_p = O_p -s_p({\mathcal C}) + s_{p-1}({\mathcal C}).$$
Since the necessary and sufficient condition to apply the rule on
position $j$ of $b$ is $b_j - b_{j+1} \geq 1$, we are going to focus
on the difference $a_j - a_{j+1}$.
$$a_j-a_{j+1}= O_j -s_j({\mathcal C}) + s_{j-1}({\mathcal C})
-(O_{j+1} -s_{j+1}({\mathcal C}) + s_j({\mathcal C}))$$
$$= O_j-O_{j+1} - 2s_j({\mathcal C}) + s_{j-1}({\mathcal
  C})+s_{j+1}({\mathcal C})$$
$$\geq O_j-O_{j+1} - 2s_j({\mathcal D}) + s_{j-1}({\mathcal
  D})+s_{j+1}({\mathcal D})$$
$$= b_j - b_{j+1} \geq 1,$$
which proves the lemma.
\qed  \end{proof}

We can now establish the following result, which states that the shot
vector associated to a sequence of transitions only depends on the
initial and final configurations.

\begin{proposition}
  Let $a$ and $b$ be two non dual configurations such that $b
  <_{gc} a$. Then all the sequences of transitions from $a$ to $b$
  have the same shot vector and so have the same length. 
\label{proposition:inde}
\end{proposition}
\begin{proof}  

Let ${\mathcal C}$ and ${\mathcal D}$ be two sequences of transitions
from $a$ to $b$~:
$${\mathcal C} : a \rightarrow c^1 \rightarrow c^2 \rightarrow \ldots
\rightarrow c^u \rightarrow b,$$
$${\mathcal D} : a \rightarrow d^1 \rightarrow d^2 \rightarrow \ldots
\rightarrow d^v \rightarrow b.$$
Let us recall that:
$$s({\mathcal C}) = s_p({\mathcal C}) *(1,\ldots,1) + d(a,b),$$
$$s({\mathcal D}) = s_p({\mathcal D}) *(1,\ldots,1) + d(a,b).$$
Assume that $s_p({\mathcal D}) > s_p({\mathcal C})$. We have 
$s({\mathcal D}) > s({\mathcal C})$. We are going to show that there
exists a path of positive length from $b$ to $b$, which is a 
contradiction. For that, we are going to build step by step a sequence
of transitions from $b$ to $b$: $b \rightarrow e_{1} \rightarrow
\ldots \rightarrow e_{l} \rightarrow b$. For $i\leq v$ let us denote by
${\mathcal D}^i$ the following sequence:
$${\mathcal D} : a \rightarrow d^1 \rightarrow d^2 \rightarrow \ldots
\rightarrow d^i.$$
There exists a first index $i$ such that
$s({\mathcal D}^i) \not\leq s({\mathcal C})$, which implies that there
exists $j$ such that $s_j({\mathcal D}^i) > s_j({\mathcal C})$ and
$\forall j' \neq j, \quad s_{j'}({\mathcal D}^i) \leq s_{j'}({\mathcal
  C})$. Since $i$ is the first index having this property,
we have $s({\mathcal D}^{i-1}) \leq s({\mathcal C})$, so
$s_j({\mathcal D}^{i-1}) = s_j({\mathcal C})$ and $s_j({\mathcal
  D}^{i}) = s_j({\mathcal C}) +1$. Since $d^{i-1}$ and $a$ satisfy the
conditions of Lemma~\ref{lemmes:pos}, we can apply the rule on
position $j$ to $b$ to obtain a new configuration denoted by $e^1$.
Let ${\mathcal E}^1$ be the following sequence of transitions:
$$a \rightarrow c^1 \rightarrow c^2 \rightarrow \ldots \rightarrow c^u
\rightarrow b \rightarrow e^1.$$
We then have $s({\mathcal D}^i) \leq
s({\mathcal E}^1) \leq s({\mathcal D})$. By iterating this process, we
can define $e^2, e^3, ...$ Since
$|s({\mathcal E}^i)| - |s({\mathcal C})| = i$ and $s({\mathcal C}) \leq
s({\mathcal E}^i) \leq s({\mathcal D})$, after $l= |s({\mathcal D})|
- |s({\mathcal C})|$ steps, we will obtain $e^l \rightarrow b$,
which is the contradiction. Since the case where $s_p({\mathcal
  D}) < s_p({\mathcal C})$ is similar, $s_p({\mathcal D})
=s_p({\mathcal C})$ and therefore $s({\mathcal D}) = s({\mathcal
  C})$, which achieves the proof.
\qed  \end{proof}

Using this result, we can define the shot vector $s(a,b)$ for any couple
 of non dual configurations $a$ and $b$ such that $b <_{gc} a$
as  being equal to the shot vector of any sequence of
transitions from $a$ to $b$. This shot vector will be very useful in
understanding more precisely the structure and properties of the game.

\section{Lattice structure of the Game of Cards}

We dispose now of the tools we need for studying the structure of
the set of all configurations that can be obtained from a given
initial configuration $O=(O_1,\ldots,O_p)$. Let us denote by $GC(O)$
the set of all non dual configurations reachable from $O$ to which we
add $\bot$ as unique minimal element if $q > 0$ ($GC(O)$ is then the
restriction of ${\mathcal R(G)}$ to the configurations reachable from
$O$).  We are going to study the order $(GC(O), <_{gc})$ (in the
following, for simplicity reasons, $GC(O)$ will both denote the set
itself and the associated partial order). Let us first characterize
this order.

\begin{theorem}
Let $a$ and $b$ be two non dual configurations of $GC(O)$, then
$$a >_{gc} b \Leftrightarrow s(O,a) < s(O,b).$$
\label{theo:carac}
\end{theorem}
\begin{proof}   

In order to show that
$$s(O,a) < s(O,b) \Rightarrow a >_{gc} b,$$
we can consider two sequences of transitions, one from $O$ to
$a$ and the other from $O$ to $b$, and then make a similar proof as the one used
in Proposition~\ref{proposition:inde}.

On the other hand, let $a$ and $b$ be two non dual 
configurations of $GC(O)$ such that $b <_{gc} a$. Let ${\mathcal E}$ be a
sequence of transitions from $a$ to $b$. The sequence 
${\mathcal D}$ built by concatenating the sequence ${\mathcal C}$ from
$O$ to $a$ and the sequence ${\mathcal E}$ is clearly a sequence from
$O$ to $b$, and so we obtain:
$$s(O,b) = s({\mathcal D}) = s({\mathcal C})+ s({\mathcal E})=
s(O,a)+s(a,b) >s(O,a).$$
\qed  \end{proof}

We can now establish the underlying lattice structure of the Game of
Cards.

\begin{theorem}
  $GC(O)$ is a lattice. If $a$ and $b$ are two elements of $GC(O)$ 
  different from $\bot$, then the configuration $c$ such that $s(O,c) =
  ((max(s_i(O,a), s_i(O,b)))_{i \in [1,p]}$ is reachable from O;
  if $c$ is not dual, then $inf_{gc}(a,b) = c$, otherwise
  $inf_{gc}(a,b) = \bot$.
\label{theo:treillis} 
\end{theorem}
\begin{proof}  

Let us assume that $s(O,a)$ and $s(O,b)$ are incomparable (otherwise $a$
and $b$ are comparable and the result is obvious). We are first going
to prove that $c$ is reachable from $a$. For that, we just have to
find a configuration $a'$ such that $a
\rightarrow a'$ and $s(O,a') \leq s(O,c)$. Let 
$$O \rightarrow d_1 \rightarrow d^2 \rightarrow \ldots \rightarrow d^v
\rightarrow b$$
be a sequence of transitions from $O$ to $b$ and let $i$ be the first
index such that $s(O,d^i) \leq s(O,a)$ and $s(O,d^{i+1}) \not
\leq s(O,a)$. Let us call $j$ the position on which the transition
is applied on $d^i$. Clearly $s_j (O,d^i) \leq s_j(O,a)$ and
$s_j(O,d^{i+1}) > s_j (O,a)$. Since $a$ and $d^i$ verify the condition
of Lemma~\ref{lemmes:pos}, we can apply the transition on position $j$
of $a$ to obtain a new configuration $a'$. We have $\forall j' \neq
j, \quad s_{j'}(O,a') = s_{j'}(O,a) \leq s_{j'}(O,c)$ and $s_j(O,a') =
s_j(O,d^{l+1}) \leq s_j(O,b) \leq s_j(O,c)$, which proves that $c$ is
reachable from $a$. The proof is similar for $b$. This implies that
$c$ is reachable from $O$, and by definition $c$ is the greatest
configuration smaller than both $a$ and $b$. If $c$ is not dual, it is
the greatest lower bound of $a$ et $b$, and if $c$
is dual, $\bot$ is then this lower bound. Since $GC(O)$ also has a greatest
element, it is a lattice, which ends the proof.
\qed  \end{proof}

\section{Convergence time}

We are now going to focus on the time needed either to arrive to the
unique stable configuration or to twice through to the same
configuration. For that, we are going to use
Proposition~\ref{proposition:inde} which states that all the sequences
between two non dual configurations have the same length. So we are
going to build a particular path between a given initial configuration
and either the fixed point if $q = 0$ or a particular dual
configuration if $q > 0$.

Let us first study the case $q=0$ where all the sequence converge to a
unique fixed point $(k,\ldots,k)$ denoted by $P$.

\begin{theorem}
\label{theo:inactifq=0}
  If $q=0$, then from any initial state $O$, there always exists a
  player that never can give a card to its neighbor. 
\end{theorem} 
\begin{proof}   

Assume that $O \neq P$.  Let $i$ be such that $d_i(O,P) = min_{1 \leq
  j \leq p}(O,P)$. We are first going to show step by step that there
exists a path from $O$ to $P$ in which player $i$ never plays. Let $m$
be such that $O_m$ is maximal among the $(O_j)_{1 \leq j \leq p}$ and
such that $O_{m+1} < O_m$.  Since $O \neq P$, such $m$ exists and $O_m
> k$. We have $d_m(O,P) = d_{m-1}(O,P) + O_m - k \geq d_{m-1}(O,P)+1$,
so $m \neq i$ and $d_m(O,P) \geq d_i(O,P)+1$. Let $a$ be the
configuration obtained from $O$ by applying the rule on position $m$.
We are going to show that $a$ is such that $d_i(a,P) = min_{1 \leq j
  \leq p}d_j(a,P)$.

If $m=p$, then $d_m(a,P)
=d_m(O,P) =0$ and for all $j \neq p$, $d_j(a,P) = d_j(O,P)+1$. If  $m
\neq p$, then $d_m(a,P) =d_m(O,P) -1$ and for all $j \neq p$,
$d_j(a,P) = d_j(O,P)$. In the two cases $d_m(O,P) \geq
d_i(O,P)+1$ and so $d_m(a,P) \geq d_i(O,P)$, which
$d_i(a,P) = min_{1 \leq j \leq p}d_j(a,P)$.

By iterating the process, we arrive to the fixed point $P$ by a path
with no transition in position $i$. Therefore $s_i(O,P) = 0$ and then
for all configuration $a$ between $O$ and $P$, $s_i(O,
a) = 0$ (for a given $i$, $s_i$ can only increase when following a
path).
\qed  \end{proof}

We obtain now the following corollary which directly comes from the
previous proof:

\begin{corollary}
\label{corollary:tempsq=0}
If $q = 0$ and if the initial configuration is $O$, the game ends after
  $t$ steps, with
$$t= p*(- min_{1 \leq i \leq p} d_i(O,P)) + \sum_{i=1}^{i=p}
d_i(O,P).$$
\end{corollary}

Let us finish by considering the case $q>0$. Here, it is more
difficult to give an exact formula of the time necessary to arrive on
a dual configuration, since all the path leading to such a
configuration may not have the same length. However, it is possible to
give an upper bound to this time. Let us consider a particular dual
configuration $P= \underbrace{k+1,\ldots,k+1}_{q} ,
\underbrace{k,\ldots,k}_{p-q}$. We obtain the following result (the
proof is identical to the proof of Theorem~\ref{theo:inactifq=0}).
\begin{theorem}
\label{theo:inactifq>0}
  Assume $q > 0$ and let $O$ be an arbitrary configuration. Let $i$
  be the first index such that $d_i(O,P)$ is a minimal component of
  $d(O,P)$. Then all the paths from $O$ to $P$ without circuits have
  the same shot vector where $s_i(O,P) =0$, that is
  in which player $i$ never plays. This shot vector is given by the
  following formula:
$$s(O,P)= - (min_{1 \leq i \leq p} d_i(O,P) *(1,\ldots,1)) + d(O,P).$$
Moreover, the time to reach $P$ from $O$ is equal to 
$$t= p*(- min_{1 \leq i \leq p} d_i(O,P)) + \sum_{i=1}^{i=p} d_i(O,P).$$
\end{theorem} 

\begin{corollary}
\label{corollary:tempsboucle}
If $q > 0$ and if the game starts from a given configuration $O$, then
the configuration obtained after 
$$p*(- min_{1 \leq i \leq p} d_i(O,P)) + \sum_{i=1}^{i=p} d_i(O,P) +
q(p - q) + 1$$
steps has been obtained earlier.
\end{corollary}
\begin{proof}  
Let us consider the dominance ordering on dual configurations, in
which a configuration $a$ is greater or equal to a configuration
$b$ if and only if $ \forall j \in [1,n]$, 
 $\sum_{i=1}^{i=j}a_{i} \geq
\sum_{i=1}^{i=j}b_{i}$. The
greatest element of this order is clearly $P$ and the maximal length
of a chain in this order is $q(p - q)$. Let $a$ and $b$ be two dual
configurations such that $a$ covers $b$ in this order. It is clearly
  possible to go from $a$ to $b$ by a transition, so the
covering relations are a subset of the set of transitions between dual
configurations. Since the dual configurations are the unique non
trivial strongly connected component of $\mathcal G$,
it is clear that the maximal length of a path between two dual
configurations in $\mathcal G$ is $q(p - q) + 1$, which proves the
corollary. 
\qed  \end{proof}

\bibliographystyle{plain}
\bibliography{Spiles}
\end{document}